\documentclass{mem}
\usepackage{natbib}
\usepackage{txfonts}
\usepackage{balance}
\usepackage{graphicx}
\usepackage[a4paper]{hyperref}
\idline{0}{1}
\def\LSXI{\mbox{LS~III~+46~11}}
\def\LSXII{\mbox{LS~III~+46~12}}
\def\BXC{\mbox{Berkeley~90}}
\def\teff{\mbox{$T_{\rm eff}$}}

\def\ebv{\mbox{$E(4405-5495)$}}
\def\rv{\mbox{$R_{5495}$}}

\def\AV{\mbox{$A_V$}}
\def\mum1{\mbox{$\mu$m$^{-1}$}}
\def\chir{\mbox{$\chi^2_{\rm red}$}}
\newcommand{\EW}[1]{\mbox{$W$({#1})}}
\newcommand{\KI}[1]{\mbox{K\,{\sc i}~$\lambda${#1}}}
\begin{document}

\title{The ISM in O-star spectroscopic surveys: GOSSS, OWN, IACOB, NoMaDS, and CAF\'E-BEANS}

%\subtitle{}

\author{J. Ma{\'\i}z Apell\'aniz}

%\offprints{P. Bonifacio}

\institute{
Centro de Astrobiolog{\'\i}a, CSIC-INTA, ESAC campus, Madrid, Spain \hfill $\;$ \linebreak
\email{jmaiz@cab.inta-csic.es}
}

\authorrunning{Ma{\'\i}z Apell\'aniz}

\titlerunning{The ISM in O-star spectroscopic surveys}
 
\abstract{
I present results on the interstellar medium towards the O stars observed in five optical spectroscopic surveys: GOSSS, OWN, IACOB, 
NoMaDS, and CAF\'E-BEANS. I have measured both the amount [\ebv] and type [\rv] of extinction towards several hundreds of Galactic 
O stars and verified that the \citet{Maizetal14a} family of extinction laws provides a significantly better fit to optical+NIR Galactic 
extinction than either the \citet{Cardetal89} or the \citet{Fitz99} families. \rv\ values are concentrated between 3.0 and 3.5 but for low 
values of \ebv\ there is a significant population with larger \rv\ associated with H\,{\sc ii} regions. I have also measured different DIBs 
and I have found that \EW{5797}/\EW{5780} is anticorrelated with \rv, a sign that extreme $\zeta$ clouds are characterized not only by low 
ionization environments (as opposed to $\sigma$ clouds) but also by having a larger fraction of small dust grains. The equivalent width of 
the ``Gaia DIB'' (8621~\AA) is strongly correlated with \ebv, as expected, and its behavior appears to be more $\sigma$-like than 
$\zeta$-like. We have also started analyzing some individual sightlines in detail.
}
\maketitle{}

\section{O-star spectroscopic surveys}

$\,\!$\indent There are several ongoing ground-based optical spectroscopic surveys of early-type stars (mostly of O spectral type but also
including B and WR stars) whose main goal is to study the properties of the stars. However, those surveys include large amounts of
information on the intervening ISM, which we are also analyzing \citep{Penaetal13,Maizetal14b}. 

The Galactic O-Star Spectroscopic Survey (GOSSS, \citealt{Maizetal11}) is obtaining blue-violet, $R\sim 2500$, high S/N spectroscopy of
all O-star candidates that are accessible to the survey telescopes. Results have already been published in two survey papers 
\citep{Sotaetal11a,Sotaetal14} and elsewhere (e.g. \citealt{Walbetal10a,Walbetal11}). To date, we have observed and processed 2087 stars
and several hundred more have been observed but are still unprocessed. The processed sample includes $\sim$900 O stars and $\sim$900 B
stars up to $\AV\sim 12$. The faintest stars are being observed with the 11~m GTC in the north and the 8~m Gemini in the south.

\begin{figure*}
\centerline{\includegraphics[width=0.65\linewidth]{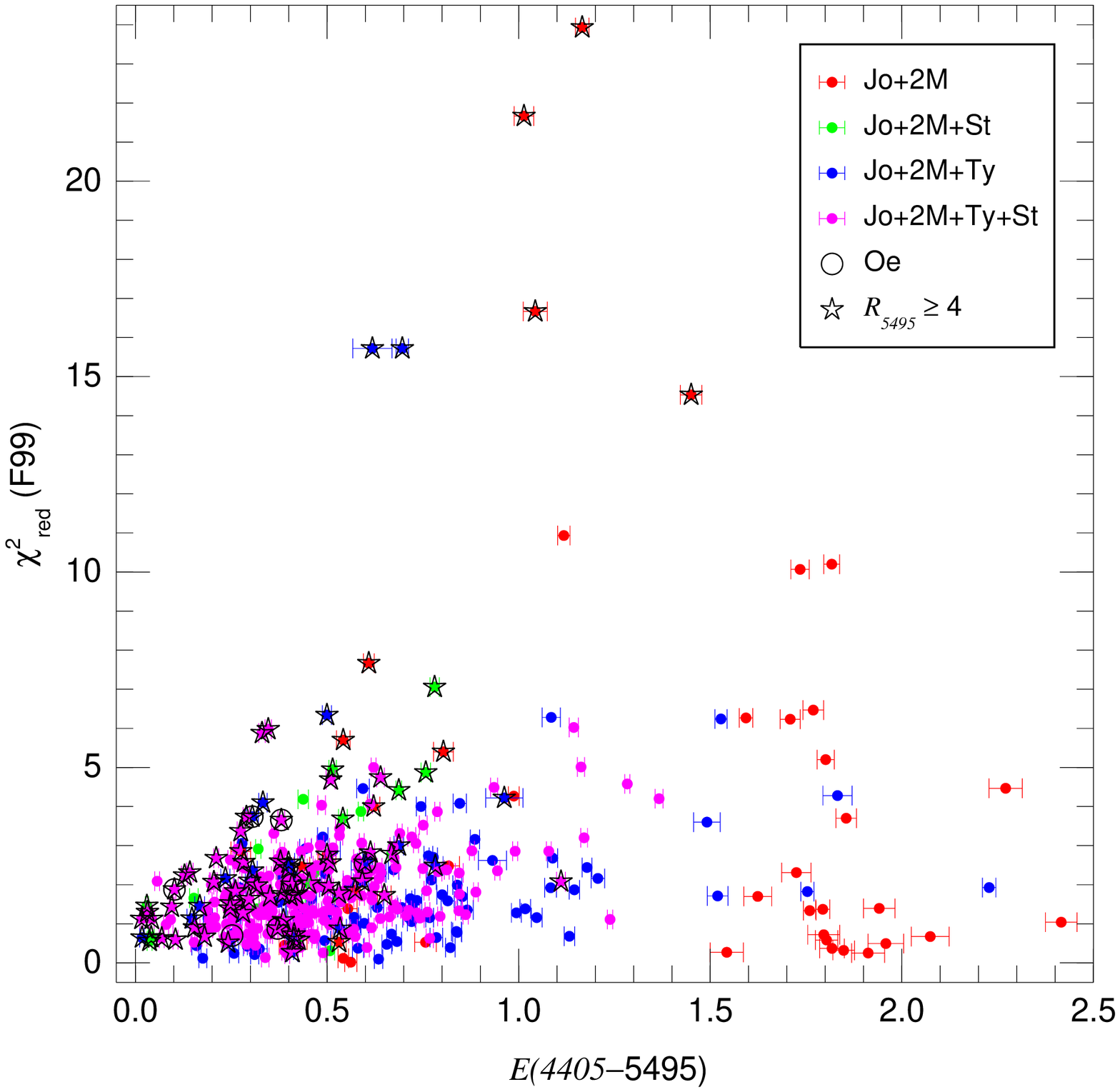} \
            \includegraphics[width=0.65\linewidth]{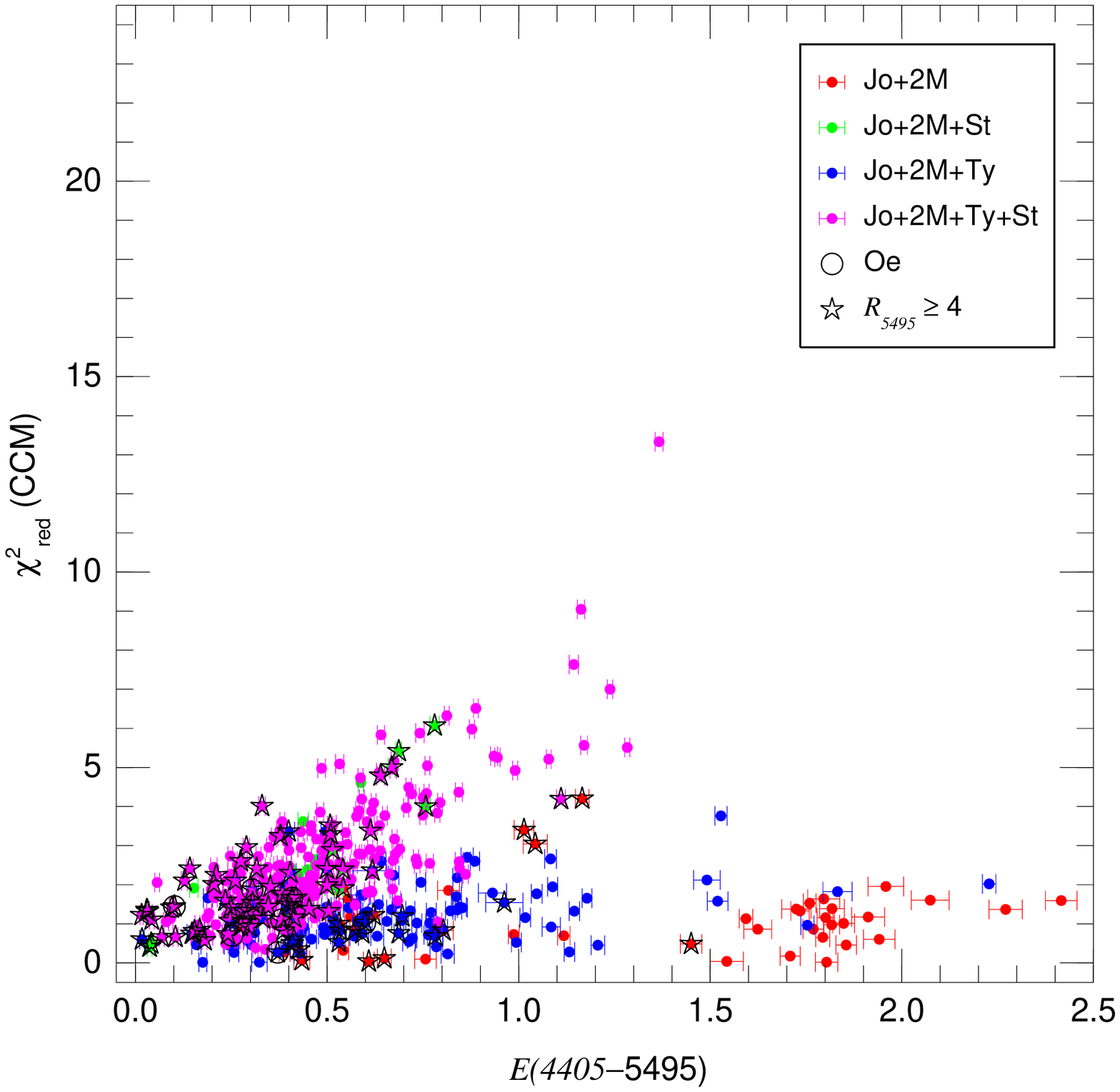}}
\centerline{\includegraphics[width=0.65\linewidth]{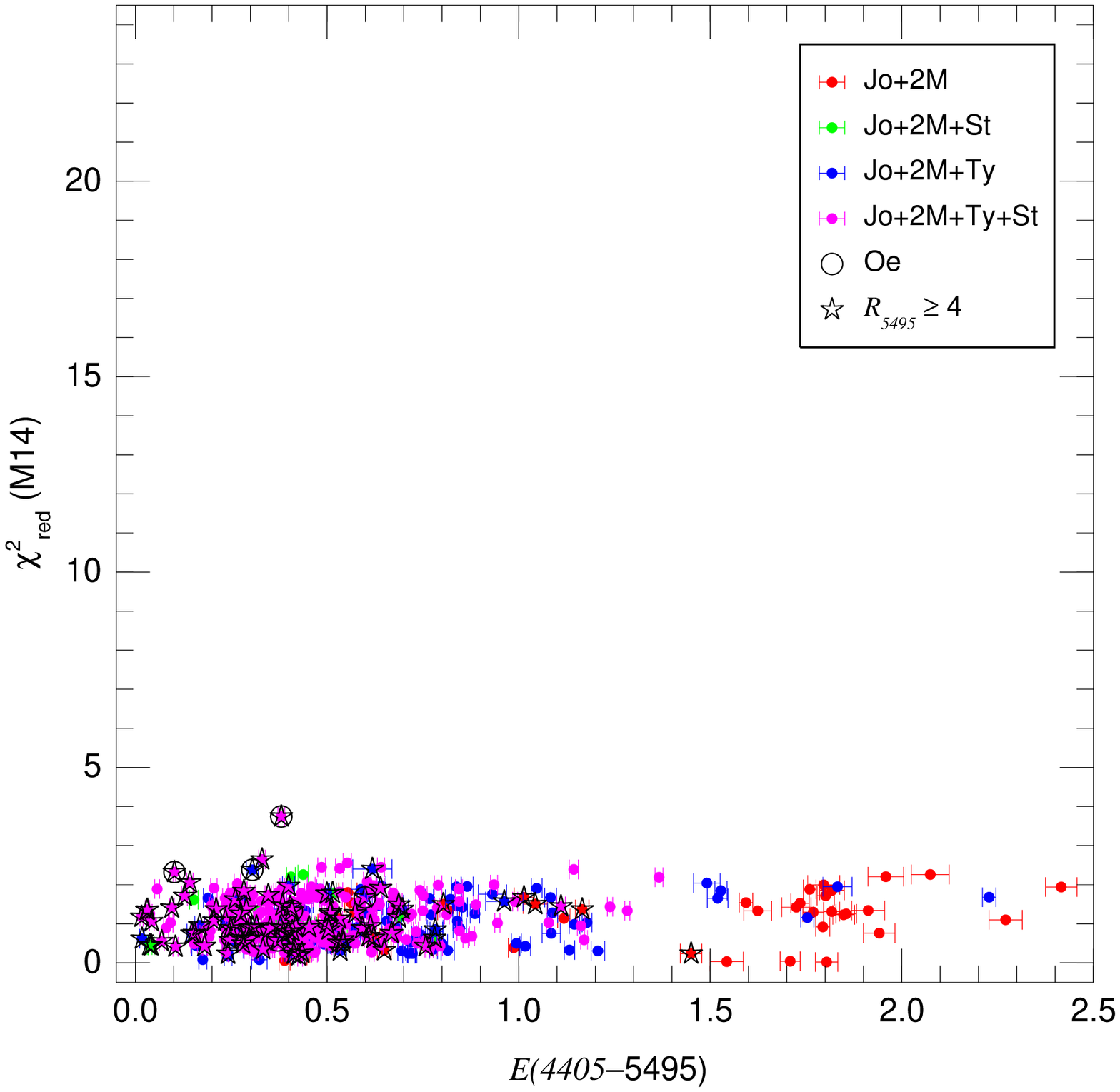}}
\caption{\footnotesize
\chir\ of the CHORIZOS fits as a function of \ebv\ for the GOSSS I+II sample using the F99 [top left], CCM [top right], and M14 [bottom]
families of extinction laws. Note that the vertical scale is the same for the three plots. Different symbols are used for stars with
different sets of photometric data (Jo = Johnson, 2M = 2MASS, Ty = Tycho-2, St = Str\"omgren). Oe stars (which are expected to have poorer
fits due to IR excesses) and stars with \rv $\ge$ 4 are marked. 
}
\label{ebv_chi2}
\end{figure*}

GOSSS is reaching a large number of stars but it is limited by its intermediate resolution, coverage of only the blue-violet region (though
some of the spectra reach longer wavelengths), and single-or-few epochs available per star. There are several surveys that complement those
limitations by obtaining high-resolution spectroscopy of a subsample of the GOSSS targets. OWN is observing the southern stars using three
different telescopes and multiple epochs with the main purpose of studying spectroscopic binarity \citep{Barbetal10}. IACOB and IACOB-sweG 
are observing northern stars with the main purpose of characterizing their properties \citep{SimDetal11c}. NoMaDS is extending IACOB 
towards fainter magnitudes using the 9~m Hobby-Eberly Telescope \citep{Maizetal12,Pelletal12}. Finally, CAF\'E-BEANS is doing multi-epoch 
spectroscopy of northern stars to study their spectroscopic binarity \citep{Neguetal15}. Combining those four high-resolution surveys, 
$\sim$450 O stars and another $\sim$450 early-type stars have been observed so far.

\section{Dust: amount and type}

\begin{figure}
\centerline{\includegraphics[width=\linewidth]{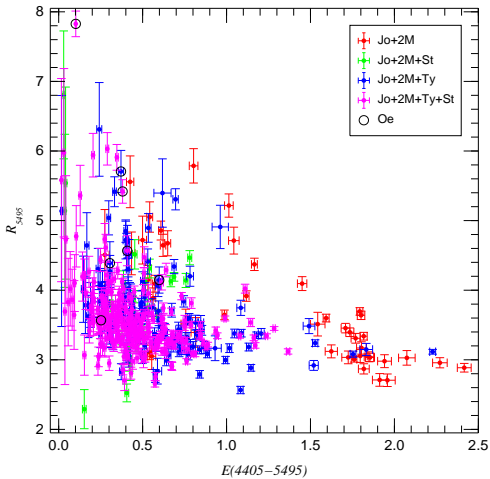}}
\caption{\footnotesize
\rv\ as a function of \ebv\ for the CHORIZOS fits for the GOSSS I+II sample using the M14 family of extinction laws. See
Fig.~\ref{ebv_chi2} for the symbol nomenclature.
}
\label{ebv_rv}
\end{figure}

\begin{figure*}
\centerline{\includegraphics[width=0.65\linewidth]{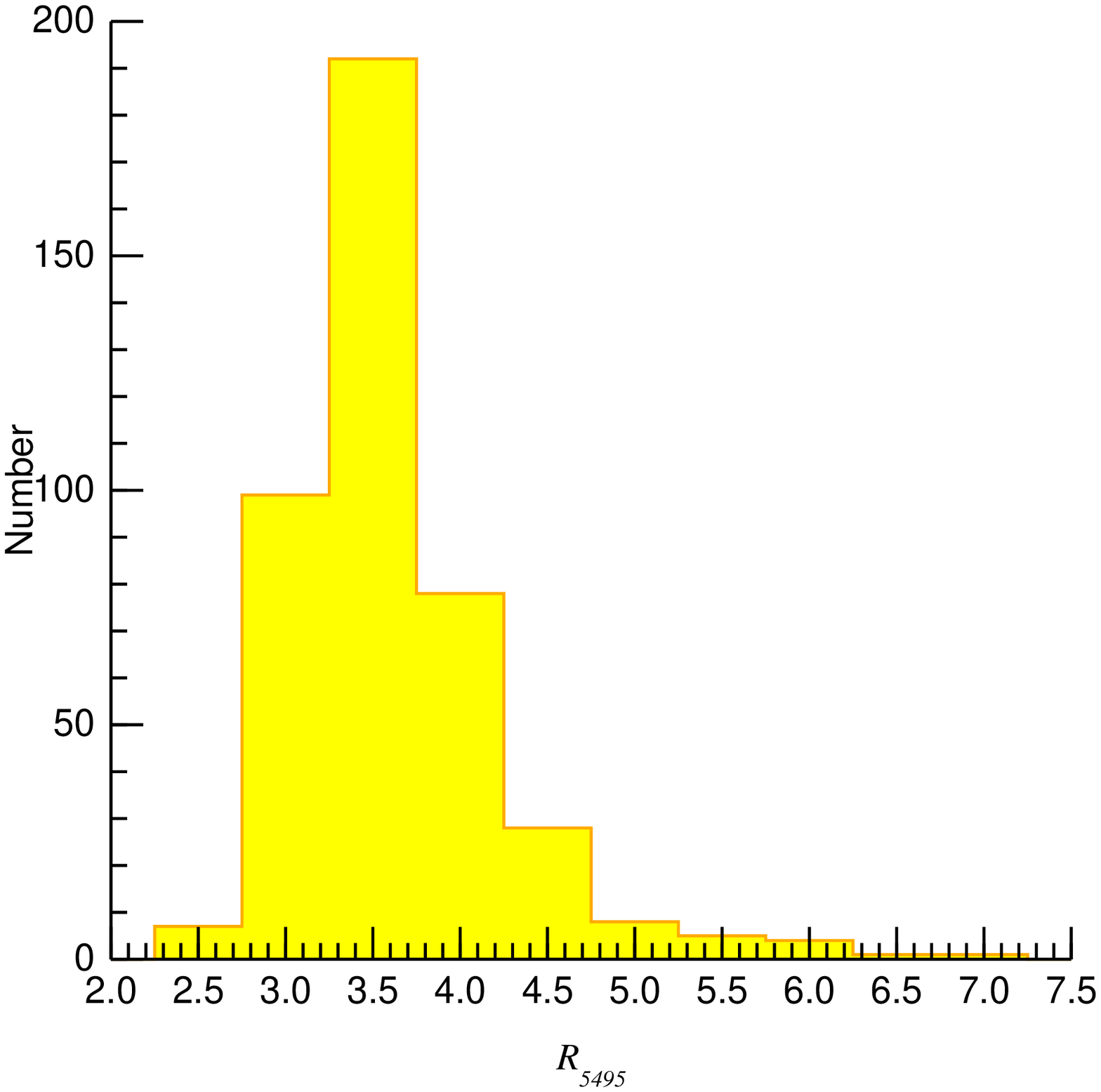} \
            \includegraphics[width=0.65\linewidth]{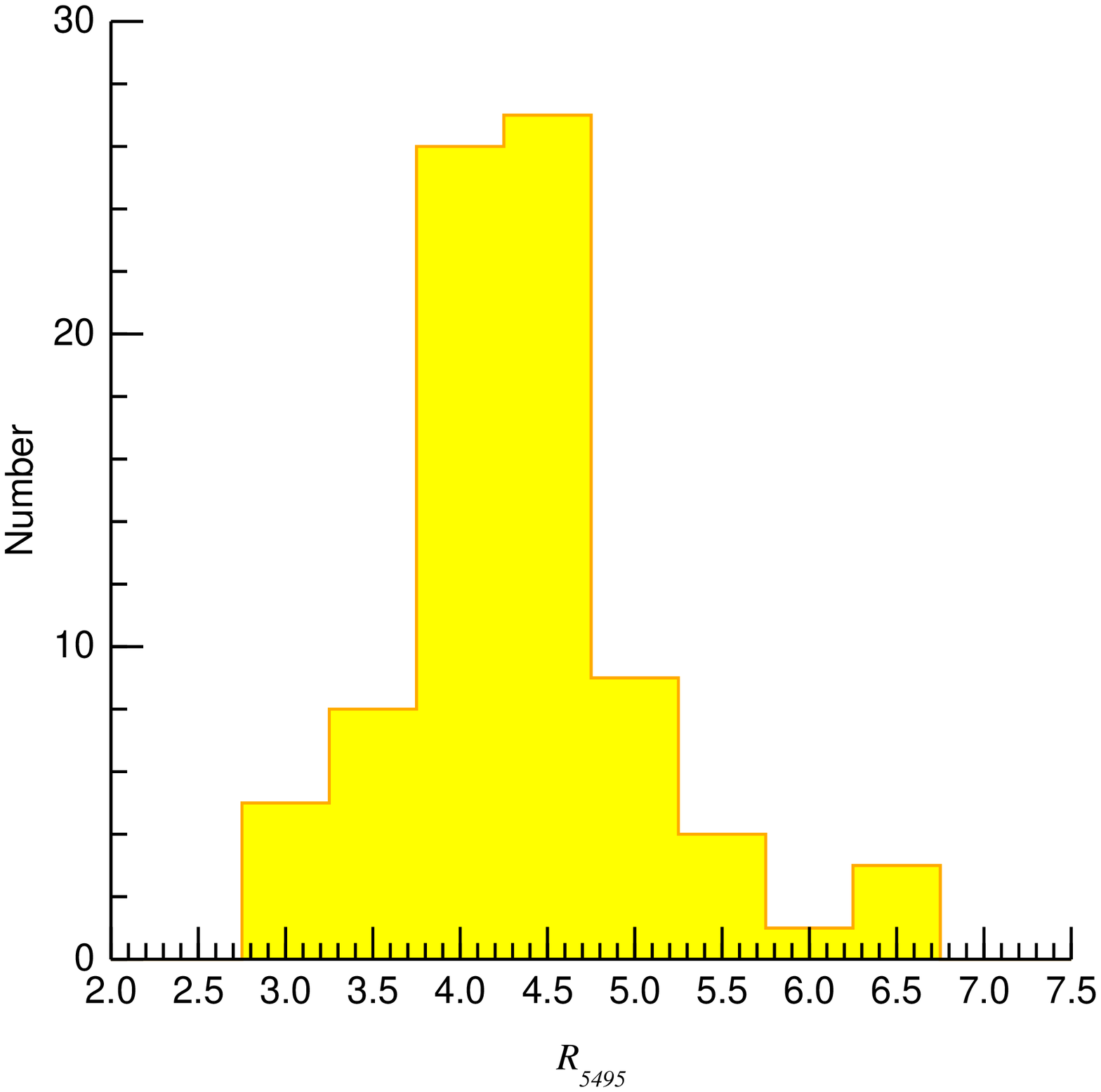}}
\caption{\footnotesize
\rv\ histograms for the (Galactic) GOSSS I+II [left] and (30 Doradus) VFTS [right] samples.
}
\label{rv_histo}
\end{figure*}

$\,\!$\indent We have recently derived a new family of optical+NIR extinction laws based on 30 Doradus data \citep{Maizetal14a}. In that
paper we used a small data set to check its validity for Galactic sightlines. Our first goal here is to extend that verification to 
the larger data set of 448 Galactic O stars of the first two GOSSS survey papers (GOSSS I+II, \citealt{Sotaetal11a,Sotaetal14}). For that
purpose we have used the Galactic O-Star Catalog (GOSC, \citealt{Maizetal04b,Sotaetal08} to collect Johnson $UBV$ and 2MASS $JHK_{\rm s}$
photometry for all of the stars in the sample (after merging some unresolved binary systems) and Tycho-2 $BV$ and Str\"omgren $uvby$ 
photometry in the cases where
they were available. The photometry was processed using CHORIZOS \citep{Maiz04c,Maiz13a} to obtain the amount [\ebv] and type [\rv] of 
extinction (see \citealt{Maiz13b} for an explanation of the choice of those quantities) in an analogous way to that used by e.g.
\citet{Ariaetal06} or \citet{Maizetal14a}. Note that this procedure can be carried out with high precision due to the use of the \teff\
derived from the GOSSS spectral types.

The CHORIZOS experiments were executed three times using the F99 \citep{Fitz99}, CCM \citep{Cardetal89}, and M14 \citep{Maizetal14a} to
test which of those families of extinction laws yields better fits to Galactic extinction. The \chir\ values (expected to be $\sim 1$ in
the ideal case) are shown in Fig.~\ref{ebv_chi2}:

\begin{itemize}
 \item The F99 results yield, on average, the poorest fits. Although they are not bad for low values of \ebv\ and \rv, they worsen 
       when either of those quantities increase (they are significantly bad for $\rv > 4$). 
 \item The CCM results are better that those of F99 in terms of \chir. However, they yield poor results when Str\"omgren photometry is used 
       due to their use of a seventh degree polynomial for interpolating in wavelength (F99 and M14 use splines, see \citealt{Maiz13b}). 
 \item The M14 results are the best of the three families. Indeed, the only case with $\chir > 3$ corresponds to an Oe star, which are 
       expected to yield poor fits in some cases due to the IR excesses produced by their disks. Therefore, the M14 family is the best 
       choice for optical+NIR extinction not only in 30 Doradus but also in the Milky Way. 
\end{itemize}

After choosing the M14 family, I show its detailed CHORIZOS results in Figs.~\ref{ebv_rv}~and~\ref{rv_histo}. The latter also includes an 
\rv\ histogram for the VFTS O-star sample in 30 Doradus \citep{Evanetal11a,Maizetal14a}. The main findings are:

\begin{itemize}
 \item As previously known for many decades, the majority of sightlines have \rv\ between 3.0 and 3.5.
 \item At very low extinctions ($\ebv < 0.2$) the error bars on \rv\ are too large to yield significant results.
 \item There are few stars with $\rv < 3.0$ (sightlines with a large proportion of small dust grains) but they are a significant fraction 
       of the objects with large extinction.
 \item In the range $0.2 < \ebv < 1.2$ there is a significant fraction of stars with large values of \rv. These sightlines are depleted in
       small grains and are associated with H\,{\sc ii} regions.
 \item As previously noted, some Oe stars have a poor fit due to their IR excesses.
 \item The \rv\ histograms for the Galaxy and 30 Doradus are markedly different, with the latter showing a larger fraction of high-\rv\
       sightlines. The differences can be explained by [a] the lower values of \ebv\ and [b] the larger fraction of H\,{\sc ii} region
       sightlines in 30 Doradus.
\end{itemize}

In the future we plan to extend the analysis to the rest of the GOSSS sample, which is on average more extinguished than the one in this
work.

\section{DIBs}

\begin{figure*}
\centerline{\includegraphics[width=0.65\linewidth]{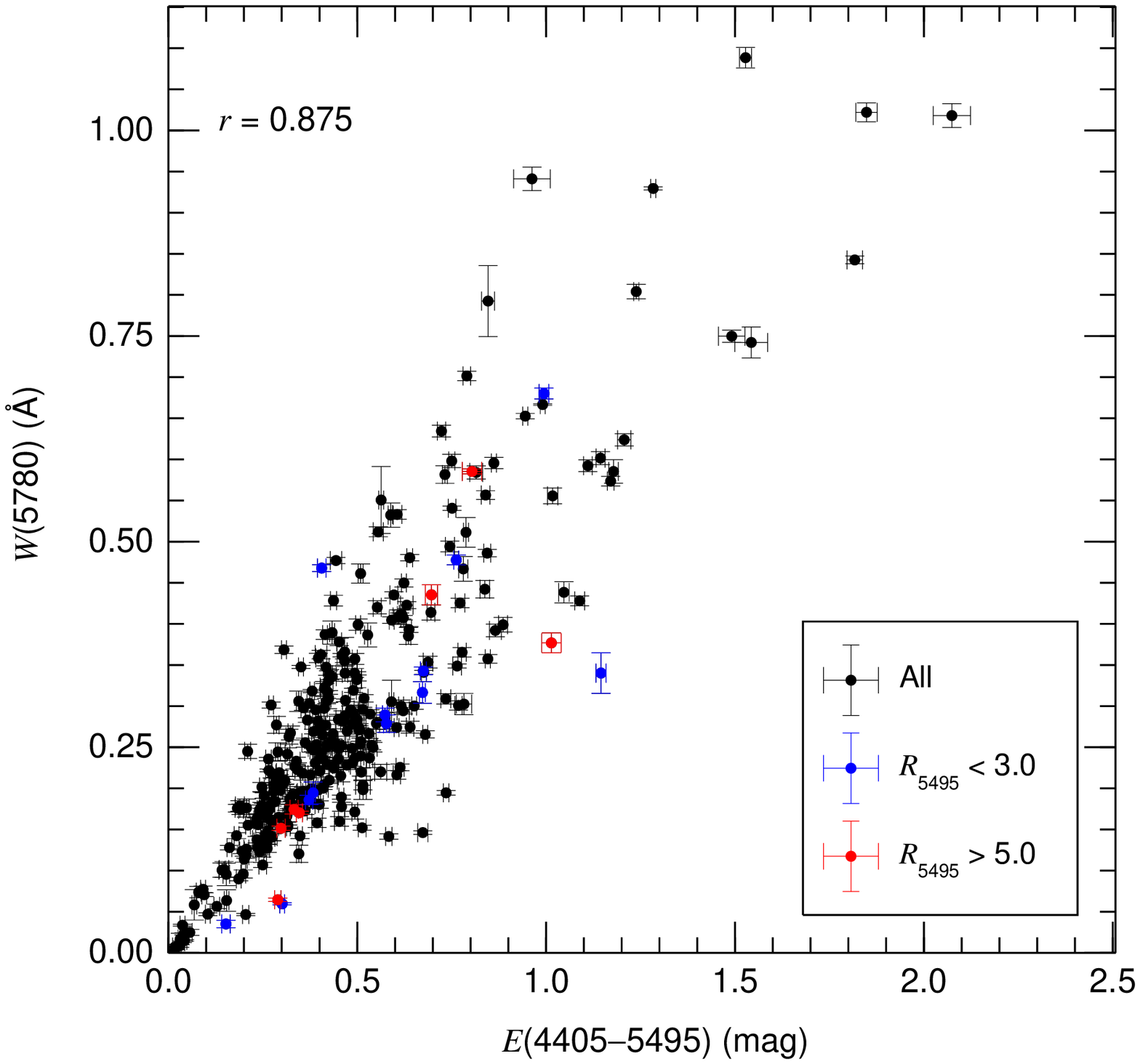} \
            \includegraphics[width=0.65\linewidth]{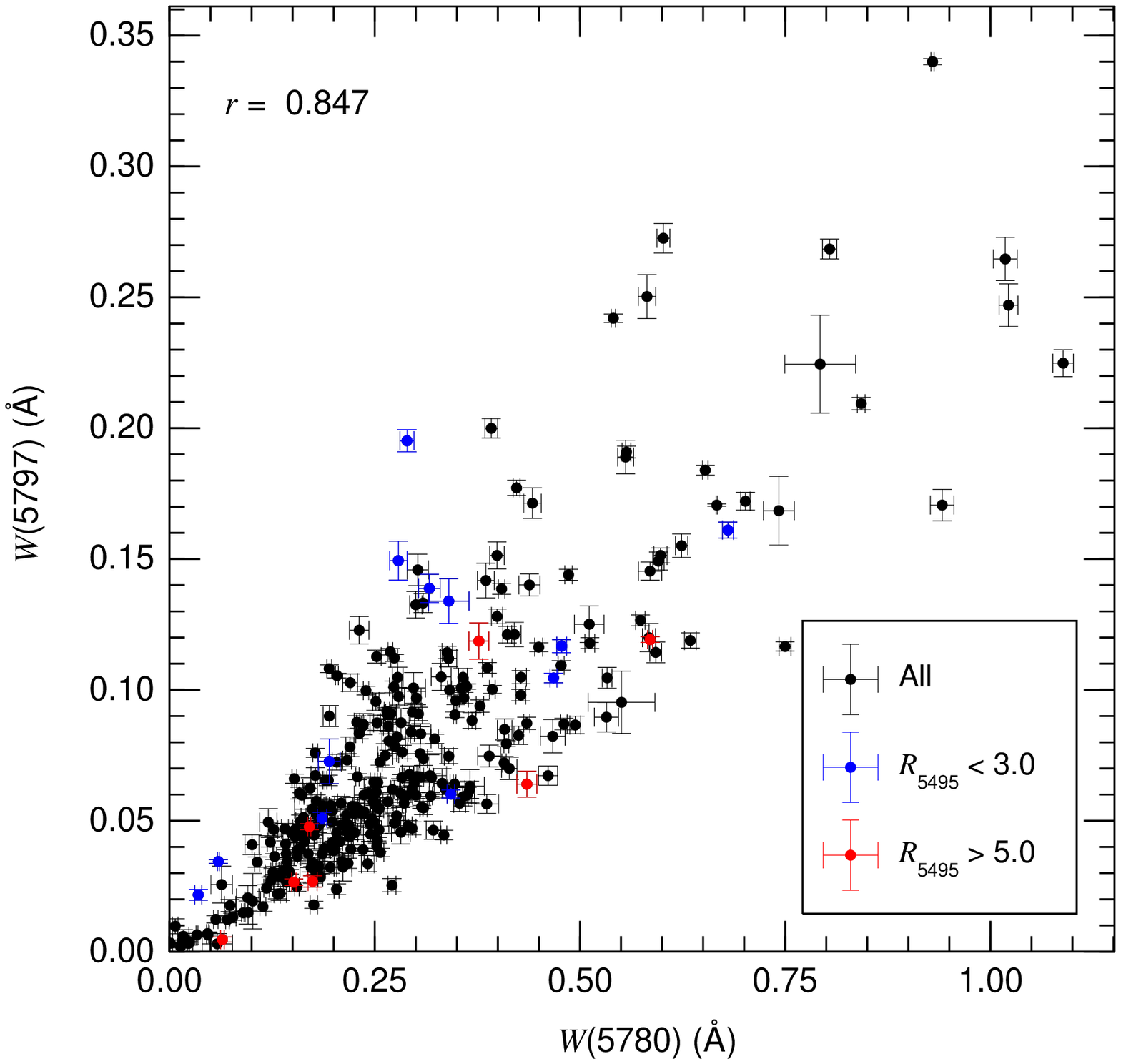}}
\caption{\footnotesize
[left] \EW{5780} as a function of \ebv\ and [right] \EW{5797} as a function of \EW{5780} for the GOSSS sample.
}
\label{ebv_dibs}
\end{figure*}

$\,\!$\indent Another line of research using these O-star surveys is the study of the Diffuse Interstellar Bands (DIBs). Our long-term 
plans include the analysis of several tens of DIBs but here I concentrate on three of them: the 5780~and~5797~DIBs (central wavelengths of 
5780.48~\AA\ and 5797.06~\AA, respectively), which are among the most studied DIBs, and the 8621~DIB (central wavelength of 8620.65~\AA), 
which is less studied but has received recent atention due to its inclusion in the wavelength region being observed by the Gaia RVS 
instrument.

We have extracted the equivalent widths ($W$) for the three DIBs above using a subset of the high-resolution spectra in the surveys, 
correcting for saturation effects in the cases with deeper absorption profiles (this is not an issue for low values of extinction but 
some of our lines are heavily extinguished). These type of studies have been done before and the originality of this ongoing work lies 
in:

\begin{itemize}
 \item The large sample in the survey, which in its final version will include over 1000 early-type star sightlines.
 \item The large dynamic range in extinction, which will eventually extend to $\AV\sim 12$ (note that the subsample presented here only
       extends to $\AV\sim 6$).
 \item The completeness down to a given magnitude (currently, $B\sim 8$, to be extended in the future) for O stars and the full-sky 
       coverage.
 \item The use of modern values of \ebv\ and \rv\ (see above) as extinction measurements.
\end{itemize}

\begin{figure}
\centerline{\includegraphics[width=\linewidth]{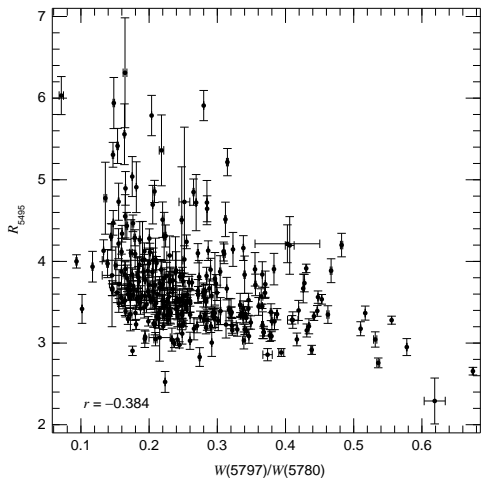}}
\caption{\footnotesize
\rv\ as a function of \EW{5797}/\EW{5780} for the GOSSS sample.
}
\label{57975780_rv}
\end{figure}

I plot in Fig.~\ref{ebv_dibs} \EW{5780} as a function of \ebv\ and \EW{5797} as a function of \EW{5780}. In both cases the two quantities
are highly correlated but the Pearson correlation coefficient (shown in both plots) is between 0.8 and 0.9, a result similar to that of
previous studies (e.g. \citealt{Frieetal11,Vosetal11,Raimetal12}). More interestingly, in the right panel we can see that sightlines with
low values of \rv\ tend to lie above the average relationship while those with high values of \rv\ tend to lie below the average
relationship. To better visualize the effect, I plot in Fig.~\ref{57975780_rv} \rv\ as a function of \EW{5797}/\EW{5780}. The Pearson
correlation coefficient in that case is $-0.384$, significant but not very large. However, the anticorrelation is clear when we examine the 
extremes of the distribution: all sightlines with $\EW{5797}/\EW{5780} > 0.50$ have $\rv < 3.5$ and all with $\rv > 5.0$ have 
$\EW{5797}/\EW{5780} < 0.35$. Putting it in another way, most sightlines concentrate in the lower left quadrant of Fig.~\ref{57975780_rv},
with some points in the lower right and upper left quadrants and none in the upper right. 

\begin{figure*}
\centerline{\includegraphics[width=0.65\linewidth]{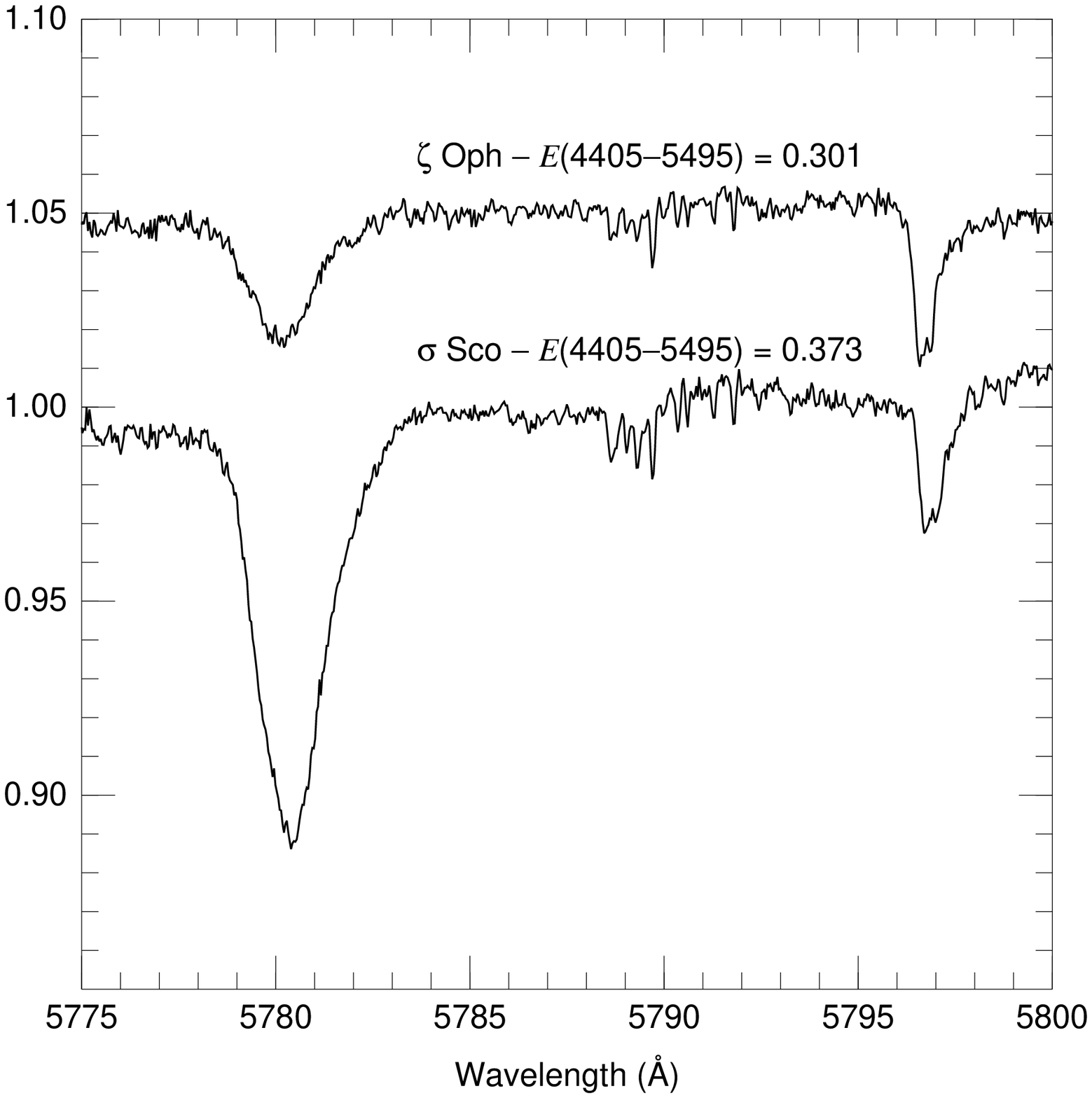} \
            \includegraphics[width=0.65\linewidth]{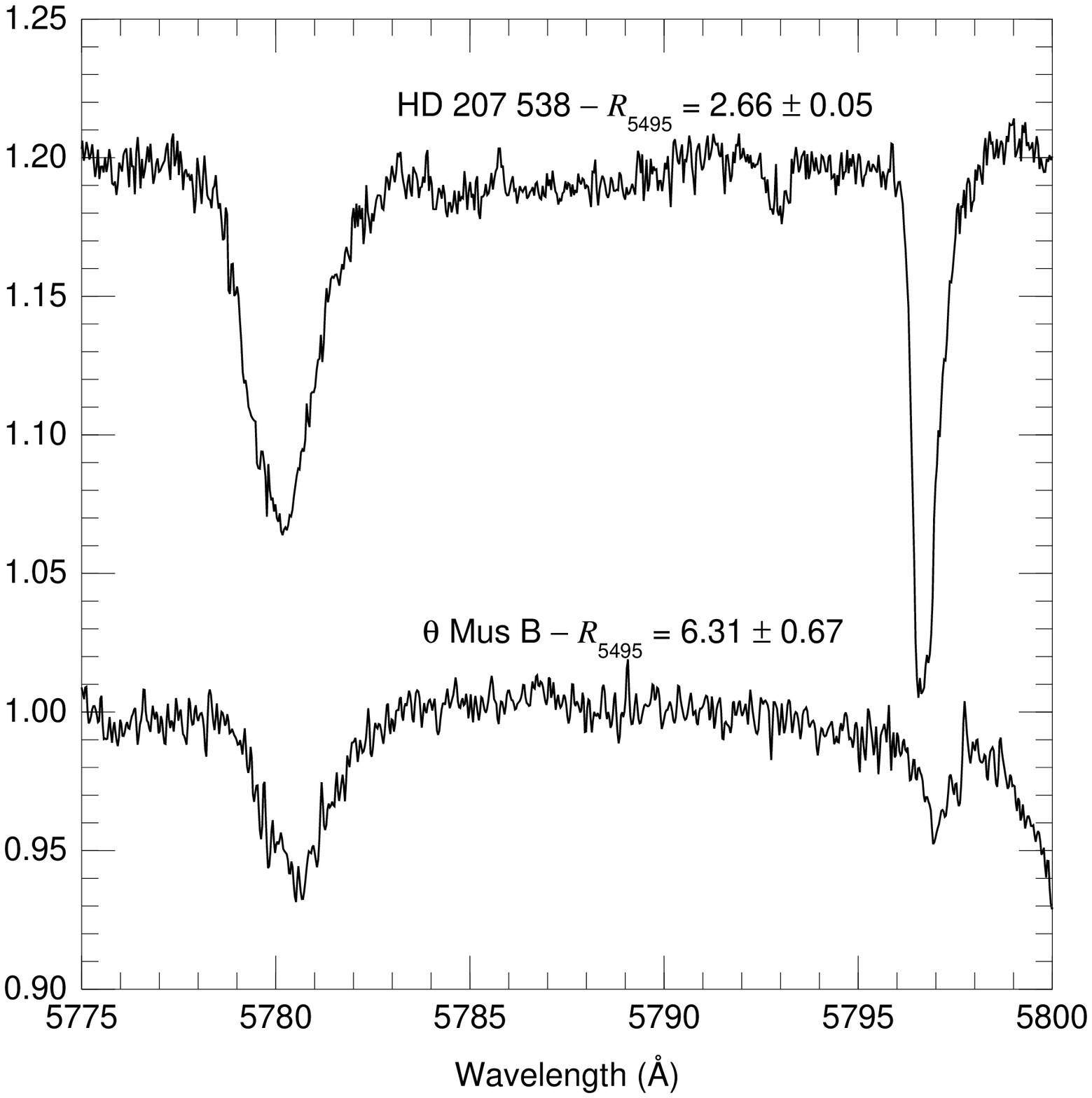}}
\caption{\footnotesize
[left] The prototype $\zeta$ ($\zeta$~Oph) and $\sigma$ ($\sigma$~Sco) sightlines for the 5780 and 5797 DIBs and 
[right] two more extreme examples, HD~207\,538 ($\zeta$) and $\theta$~Mus~B ($\sigma$).
}
\label{zeta_sigma}
\end{figure*}

How do we interpret this result? It has been known for some time \citep{Kreletal97,Coxetal05} that DIBs come in families and that the 
5780 and 5797 DIBs are representative of two types of environments. The 5797 DIB is (relatively) strong in $\zeta$~Oph, a sightline known 
to be exposed to a low UV flux, and is thus termed a ``$\zeta$ DIB''. On the other hand, the 5780 DIB is strong in $\sigma$~Sco, a 
sightline exposed to a high UV flux, and is thus termed a ``$\sigma$ DIB''. See Fig.~\ref{zeta_sigma} and notice the large difference in 
the 5780 DIB for those two sightlines, which have a similar \ebv. $\zeta$~Oph and $\sigma$~Sco are the two prototypical examples of this
phenomenon but in our sample we have found even more extreme cases such as HD~207~538 and $\theta$~Mus~B, also in Fig.~\ref{zeta_sigma}. 
What we have found is that $\zeta$ clouds (those with low UV flux) tend to have low values of \rv\ (i.e. a larger proportion of small dust 
grains) while $\sigma$ clouds (those with high UV flux) tend to have large values of \rv\ (i.e. a smaller proportion of small dust grains). 
Such a relationship between cloud type and \rv\ was proposed by \citet{Camietal97}, indicating that $\zeta$ clouds likely correspond to
cloud cores and $\sigma$ clouds to cloud skins, but here we demonstrate its existence with a large sightline sample for the first time.

The difference in grain population between denser, non-UV-exposed, $\zeta$ clouds and thinner, UV-exposed, $\sigma$ clouds can be
interpreted in terms of selective destruction of small grains in the latter. One interpretation would be that sputtering dominates over 
shattering in H\,{\sc ii} regions and other parts of the ISM exposed to UV light, as the former preferentially destroys small grains and
the latter preferentially destroys large grains \citep{Joneetal94,Andeetal11b}. However, sputtering is only expected to become efficient
above $10^5$~K for refractory grains (Fig.~25.4 in \citealt{Drai11}), so a different mechanism is required. A possibility is that the 
extreme UV radiation present in H\,{\sc ii} regions destroys the PAHs that likely constitute the small grain population \citep{Tiel08}.

\begin{figure}
\centerline{\includegraphics[width=\linewidth]{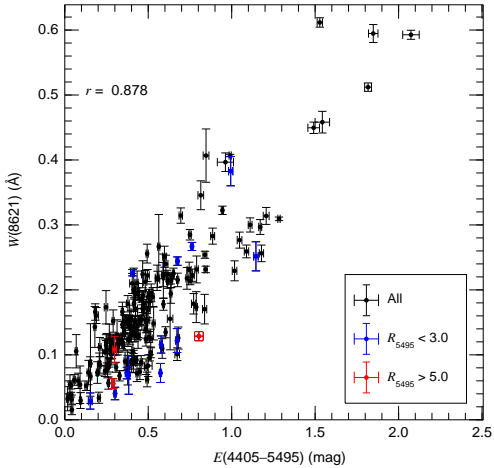}}
\caption{\footnotesize
\EW{8621} as a function of \ebv\ for the GOSSS sample.
}
\label{ebv_8621}
\end{figure}

I also show in Fig.~\ref{ebv_8621} \EW{8621} as a function of \ebv. As expected, both quantities are strongly correlated with Pearson
coefficients similar to those of other DIBs. Such measurements had been carried out before for that DIB but reaching lower values of
\ebv\ \citep{Munaetal08,Kosetal13}. In the future we will extend this study to even larger extinctions.

\section{Individual sightlines}

\begin{figure}
\centerline{\includegraphics[width=\linewidth]{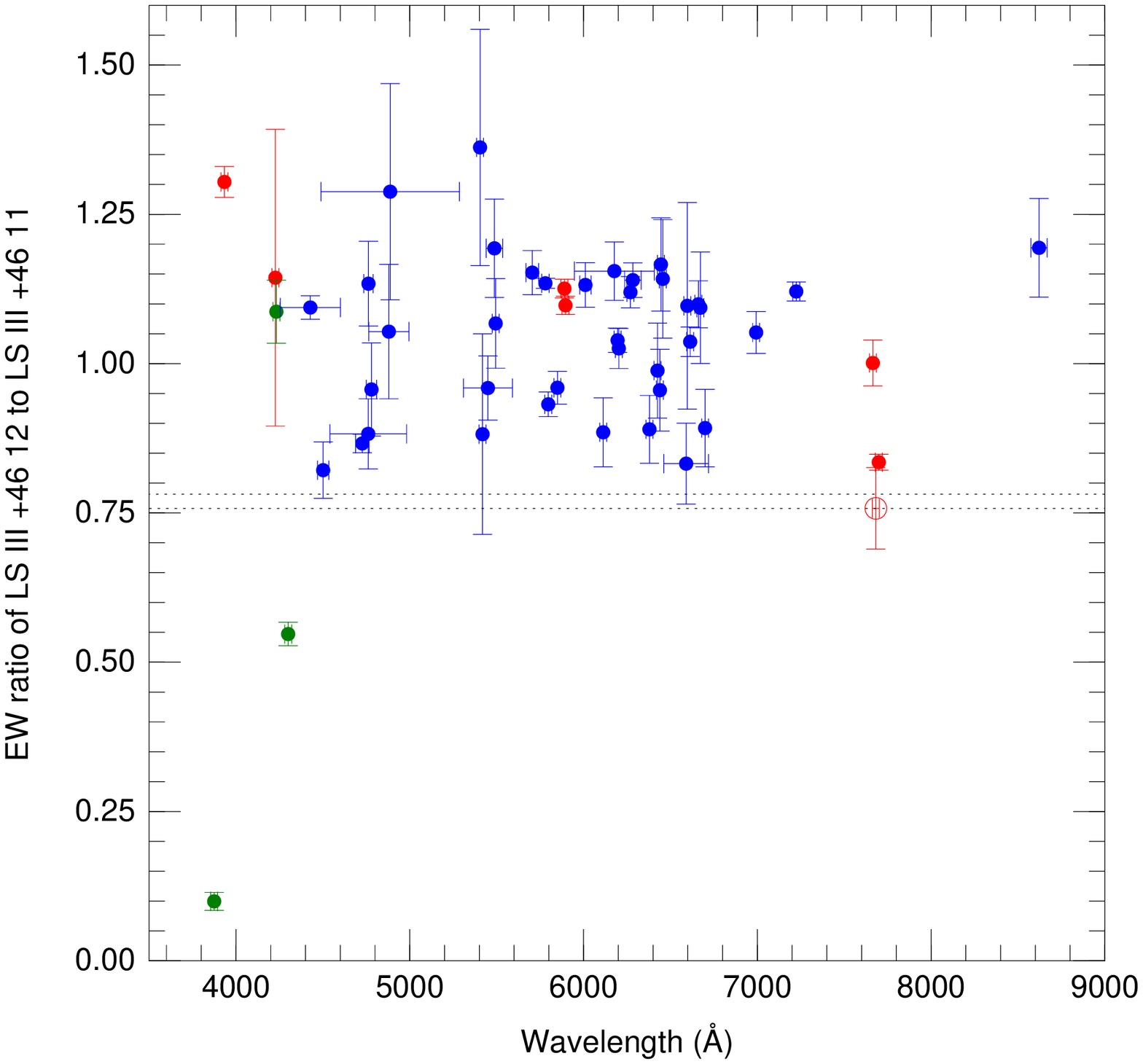}}
\caption{\footnotesize
Equivalent-width ratios between \LSXII\ and \LSXI. Red is used for atomic lines, green for molecular lines, and blue for DIBs. All points 
show the EW uncorrected for saturation (which should be a good approximation in all cases except for the intense atomic lines) except for 
\KI{7664.911+7698.974}. In that case, besides the uncorrected values we show the corrected one (as an unfilled symbol) calculated using a
kinematic decomposition. The $x$ value in each case is $\lambda_0$, the length of the horizontal error bars is proportional (with a 
threshold for the narrowest ones) to the FWHM of the line, and the vertical error bars show the uncertainty in the measurement. The two 
dotted lines are the ratios for \ebv\ and \AV.
}
\label{ewr}
\end{figure}

$\,\!$\indent A third line of work is the study of individual sightlines in detail, of which we are analyzing several. Here we present a
summary of our results on \BXC, a little-studied cluster that includes two early-O-type systems, \LSXI\ and \LSXII, separated by a short
distance \citep{Maizetal15a}. 

\begin{figure*}
\centerline{\includegraphics[width=\linewidth]{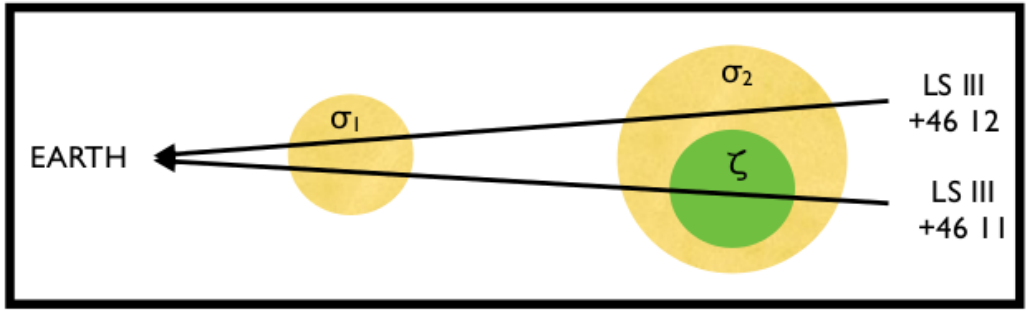}}
\caption{\footnotesize
Model for the ISM clouds present in the \LSXI\ and \LSXII\ sightlines. Cloud $\sigma_1$ is of low density, not associated with the cluster, 
and affects both sightlines similarly. Cloud $\sigma_2$ is also of low density, is the skin of the cloud associated with the cluster, and is 
longer along the \LSXII\ sightline. Cloud $\zeta$ is of high density, is the core of the cloud associated with the cluster, and affects \LSXI\
exclusively (or, at least, to a much larger degree than \LSXII).}
\label{model}
\end{figure*}

Both stars show large extinctions but \LSXI\ is significantly more extincted than \LSXII\ (by 30\%). \rv, on the other hand is very similar
for both stars and has standard values between 3.3 and 3.4. We have measured as many as seven atomic, seven molecular, and fifty DIBs for 
both stars and we have found large differences among them. We show in Fig.~\ref{ewr} the equivalent-width ratios for the lines we have
measured.
From the kinematic point of view, there are two components clearly visible in our high-resolution spectra. The weak one has a velocity
intermediate between that of the cluster and ours, appears to be of relatively low density, and affects the two stars similarly. The strong 
one has a velocity similar to that of the cluster and affects the two stars differently: for \LSXII\ it appears to be of lower density than for
\LSXI. Combining all the information, we propose a model in which three different ``clouds'' (actually, one simple cloud and one cloud with
core and skin) are located between \BXC\ and us (Fig.~\ref{model}). The reader interested in more details is referred to 
\citet{Maizetal15c}.

One interesting result of the \BXC\ analysis is that it alows us to classify a large number of DIBs in a $\sigma-\zeta$ scale, something that
was not possible before due to the lack of good data. In that sense, the 8621 DIB appears to be of $\sigma$ type i.e. it samples the diffuse,
UV-exposed ISM but not the dense, UV-shielded ISM. 

The two sightlines have different values of \EW{5797}/\EW{5780} (0.34 and 0.28 for 
\LSXI\ and \LSXII, respectively) but similar values of \rv, which is in apparent contradiction with the correlation between 
\EW{5797}/\EW{5780} and \rv\ previously discussed. However, that correlation is determined mostly by the extremes (sightlines with
large values of \rv\ and sightlines with large values of \EW{5797}/\EW{5780}), which is the reason why the Pearson coeffcient is not very
large. For intermediate values Fig.~\ref{57975780_rv} shows that it is possible to have variations in \EW{5797}/\EW{5780} without 
changing \rv. In other words, the DIB-carrier population seems to change continuously with UV-field strength/density while for \rv\ the
variation seems to happen mostly at the extremes of the distribution (in H\,{\sc ii} regions and in very dense clouds), with an intermediate
``typical'' range where \rv\ is nearly constant.

\section{Conclusions}

\begin{itemize}
 \item The M14 family of extinction laws provides better optical+NIR fits than either F99 or CCM for both the Milky Way and 30 Doradus.
 \item \rv\ and \EW{5797}/\EW{5780} are anticorrelated, indicating that extreme $\zeta$ clouds have a larger fraction of small dust grains 
       than extreme $\sigma$ clouds. However, in the intermediate region between the extremes it is possible to have clouds with the same 
       grain distribution but different proportions of $\zeta$ and $\sigma$ DIBs.
 \item The 8621 DIB is strongly correlated with \ebv\ up to $\AV\sim 6$ and the preliminary evidence for some sightlines point towards
       a $\sigma$ character.
\end{itemize}

\begin{acknowledgements}
I thank my collaborators who participate in these surveys and M. Penad\'es Ordaz, who performed an initial reduction of the DIB data. 
I also thank Bruce Draine and Jacco van Loon for useful discussions on the relationship between \rv\ and DIB types and Rodolfo Xeneize
Barb\'a for general discussion on the ISM and for ideas and material that helped me understand this topic better.
I acknowledge support from [a] the Spanish Government Ministerio de Econom{\'\i}a y Competitividad (MINECO) through grants 
AYA2010-15\,081, AYA2010-17\,631, and AYA2013-40\,611-P; [b] the Consejer{\'\i}a de Educaci{\'o}n of the Junta de Andaluc{\'\i}a through 
grant P08-TIC-4075; and [c] the George P. and Cynthia Woods Mitchell Institute for Fundamental Physics and Astronomy.
I am grateful to the Department of Physics and Astronomy at Texas A\&M University for their hospitality during some of the time this work 
was carried out. 
\end{acknowledgements}

\bibliographystyle{aa}
\bibliography{general}

\end{document}